\documentclass[%
 reprint,
 amsmath,amssymb,
 aps,
 groupedaddress,
 prl
]{revtex4-2}
\usepackage{graphicx}
\usepackage{dcolumn}
\usepackage{bm}
\usepackage[version=3]{mhchem}
\usepackage{csquotes}
\usepackage{hyperref}
\hypersetup{
    colorlinks=true,
    linkcolor=blue,
    filecolor=magenta,
    urlcolor=cyan,
}
\usepackage{color}

\begin{document}
\preprint{PRL}
\title{Information propagation in time through allosteric signaling}

\author{Tushar Modi}
\affiliation{Department of Physics, Arizona State University, Tempe, Arizona 85287, USA}

\author{S. Banu Ozkan}
\affiliation{Department of Physics, Arizona State University, Tempe, Arizona 85287, USA}
\affiliation{co-corresponding}

\author{Steve Press\'e}
\affiliation{Department of Physics, Arizona State University, Tempe, Arizona 85287, USA}
\affiliation{School of Molecular Sciences, Arizona State University, Tempe, Arizona 85287, USA}
\affiliation{co-corresponding}

\date{\today}

\begin{abstract}
Naively, one expects the information communicated by an enzyme downstream within a signaling network, in which the enzyme is embedded, to grow monotonically with the enzyme's rate of product formation. However, here we observe that this does not necessarily hold true for allosterically regulated enzymes, often observed in signaling networks. In particular, we find that the mutual information between the catalytic sites of an allosterically regulated enzyme and a receiver protein downstream in the signaling pathway depends on the transition kinetics between the different allosterically regulated states of the enzyme and their respective rates of product formation. Surprisingly, our work implies that allosteric down-regulation of an enzyme's rate of product formation may not only be used as a way to silence itself, as one would normally expect. Rather, down-regulation may also be used to increase the information communicated by this enzyme to a receiver protein downstream in a signaling pathway.
\end{abstract}

\maketitle

\section{Introduction}
Despite the fact that only a small fraction of a cell is composed of proteins (e.g., proteins constitute 17\% of E. Coli~\cite{doi:10.1002/bies.201300066}),  proteins not only mediate the key processes in cells, but also give rise to spatio-temporal signaling to control a cell's response to its local environment underlying all critical decision-making~\cite{WODAK2019566} such as a cell's development and metabolism~\cite{LINK20148}, motility~\cite{doi:10.1098/rstb.2017.0181}, immunity and cell-death (apoptosis)~\cite{Herr13042018}. 

These spatio-temporally coordinated events 
are often achieved by proteins exhibiting allostery--a phenomenon by which the binding of a molecule at one site of a protein changes the binding affinity or catalytic activity at another distant site. Several models of allostery have previously been explored to study how allosteric interactions within one enzyme (i.e., proteins which act as catalysts) modulate its activity~\cite{monod_nature_1965,cooper_allostery_1984,koshland_comparison_1966}. However, allostery goes beyond just remote modulation. It also provides ``circuit components" from which Nature builds up complex signaling networks. Here we employ an information theoretic framework to quantify how allostery propagates through protein signaling pathways in order, downstream, to modulate cellular response.

Information theory has already been proven to be useful in studying various biological phenomena such as modeling protein interaction networks~\cite{lenaerts_quantifying_2008,levine_nbit_2014, voliotis_information_2014}, evolution~\cite{adami_use_2012,de_vladar_contribution_2011,weiss_information_2000} and single molecule experiments~\cite{steve1, steve2}. Moreover, it has also been instrumental in studying the effect of allostery on the transfer of information between the allosteric regulator and the catalytic site of allosteric proteins, particularly enzymes ~\cite{komorowski_limited_2019,marzen_statistical_2013}. However, these studies primarily explored the thermodynamics of substrate binding through allostery, focusing solely on the concentration of products generated. On the other hand, the activity of the receiver protein varies in time according to the availability and on/off binding of the product produced by the enzyme. Therefore, information conveyed is encoded not only in the total concentration of the enzymatic product generated by ``sender enzymes", but also in its time varying activity.  Here, we apply information theory to quantify the role of allostery in a sender enzyme 
while it communicates with a receiver protein (a protein to which product molecules may bind).

To apply information theory, we first consider a simple master equation model capturing the essence of allostery in the arrival of products at the receiver protein. Next, this model is simplified to a two state model which is studied in a discrete time domain with the help of Hidden Markov Models (HMM)~\cite{rabiner_introduction_1986, seymore_learning_1999}. We compute the probability of observing product arrival events accounting for all possible latent allosteric states (``hidden trajectories") of the sender enzyme. We then calculate the mutual information (MI)~\cite{Cover:2006:EIT:1146355} over the joint distribution of the state of the sender enzyme and the state of receiver protein. The joint distribution  encodes both the allosterically induced state-switching transitions and the product formations by the sender enzyme's catalytic site.

Our work illustrates how allostery directly impacts the transfer of information within signaling pathways. It shows that the communication by a sender enzyme in a signaling pathway is not merely modulated by the number of products generated, but also significantly depends on the time signature of product arrivals at other receiver proteins present downstream. The analysis also suggests a broader role for allostery: a way to increase the information communicated within signaling pathways even, counter-intuitively, via down-regulation.

\section{Model}

\begin{figure}[h!]
    \centering
    \includegraphics[width=3.5in]{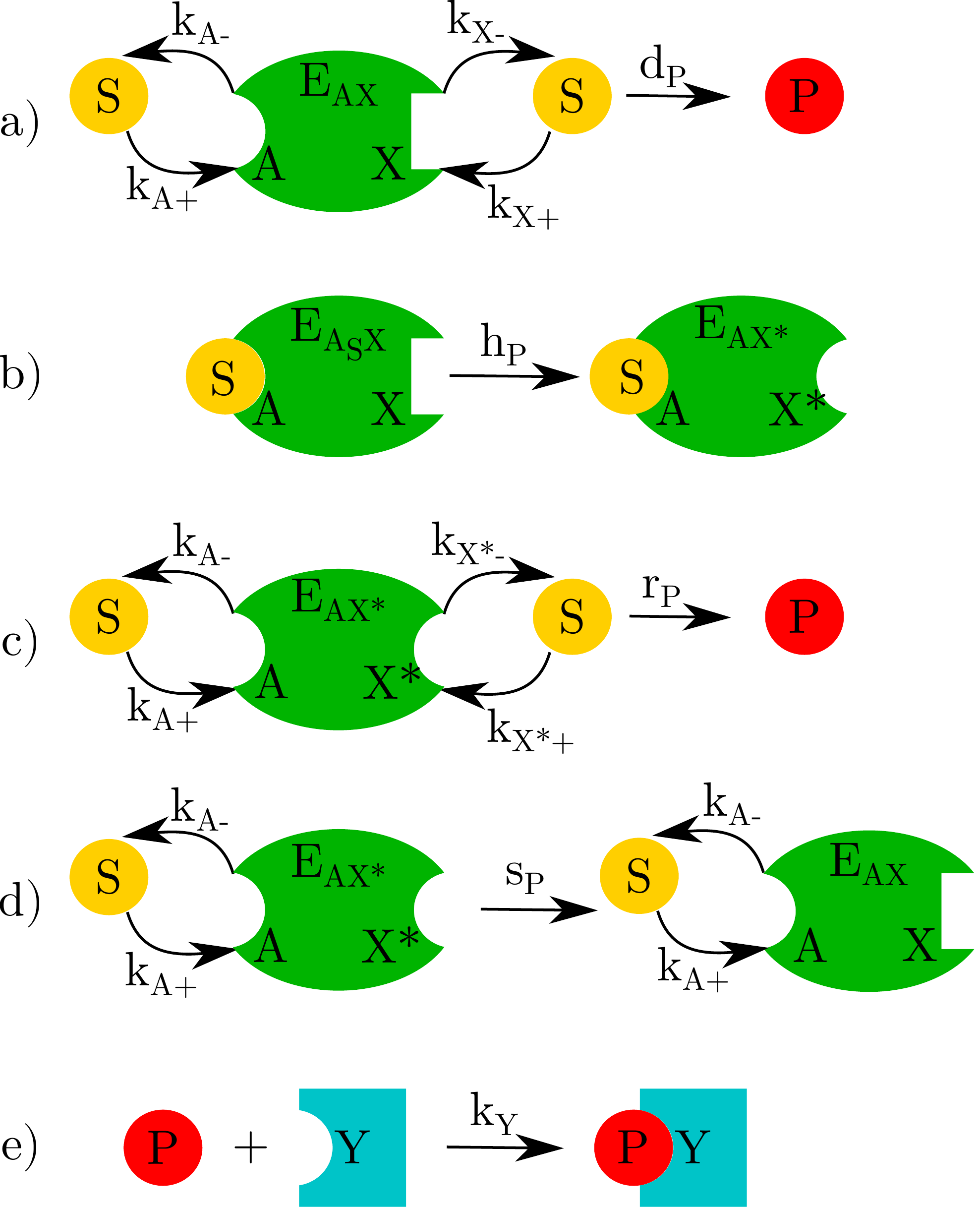}
    \caption{\label{fig:model} \textbf{Schematic diagram of the model of an allosterically regulated sender enzyme.} It shows ${E_{AX}}$, a sender enzyme, with catalytic site $X$ interacting with substrates $S$ to make product $P$ and allosteric regulator site $A$ which can interact with the substrates to change the state of the sender enzyme to ${E_{AX}}$ with catalytic site $X^{*}$. The catalytic site ${X^{*}}$ can then also interact with the substrate to make products $P$ however with different rate constants. The products generated then bind to a receiver protein $Y$ downstream in the signaling pathway.}
\end{figure}

We start with a minimal model of a sender enzyme with an allosteric regulator site (Fig.~\ref{fig:model}) inspired by the  KNF (Koshland-Nmethy-Filmer) model of protein allosteric regulation as described in Refs.~\cite{koshland_comparison_1966}. The model features a sender enzyme, $E_{AX}$ which functions as a sender of information to another protein downstream in the signaling pathway (a receiver protein) in the form of products. It has a catalytic site $X$ 
 which can bind or unbind to a substrate $S$ to make complex $E_{AX_{S}}$, with rate constants $k_{X+}$ and $k_{X-}$, respectively. In addition, the sender enzyme also contains of an allosteric regulator site $A$. The sender enzyme with the bound complex at the catalytic site subsequently degrades to generate product, $P$, with a rate constant of $d_p$ releasing its catalytic site $X$ back to its original unbound form.

When the allosteric regulator site $A$ reversibly binds to substrate, it creates a complex $E_{A_{S}X}$ with the rate constants $k_{A+}$ and $k_{A-}$ respectively, see Fig.~\ref{fig:model}a. The bound complex at the allosteric regulator can exploit the sender enzyme's network of interactions to influence the activity of the catalytic site (see Fig.~\ref{fig:model}b). This allosteric interaction occurs between a bound allosteric regulator site and unbound catalytic site with a rate constant of $h_p$. The modified catalytic site ($X^*$) now performs its functions with different activity such that its rate constants for binding and unbinding with the substrates changes to $k_{X^{*}+}$ and $k_{X^{*}-}$ respectively. Moreover, the rate constant of the degradation of the complex at $X^{*}$ with substrate to create product also changes to $r_p$, see Fig.~\ref{fig:model}c. It should be noted that, these changes are only localized to the catalytic site, whereas, the allosteric regulator does not show any changes in its dynamics of interaction, i.e., the rate constants of binding and unbinding of substrate at the allosteric regulator site does not change as the catalytic site alters its state.

Finally, the catalytic site ($X$) after being allosterically modified to ($X^*$) can relax back to its original state in a process with a rate constant of $s_p$. This process can occur regardless of the state of the allosteric regulator (i.e., whether it is bound or unbound), see Fig.~\ref{fig:model}d.

The products, $P$, produced in processes a) and c) in the Fig.~\ref{fig:model} can interact with the receiver protein $Y$ downstream in the signaling pathway with a rate constant of $k_{Y}$. This protein acts as a receiver for the signal generated by the sender enzyme in the signaling pathway in the form of products. During the interaction of product $P$ with the receiver $Y$, a binding event signifies a successful transfer of signal from the allosterically regulated sender enzyme, $E$ to the receiver protein, $Y$.

\begin{figure}[h!]
\includegraphics[width=3.5in]{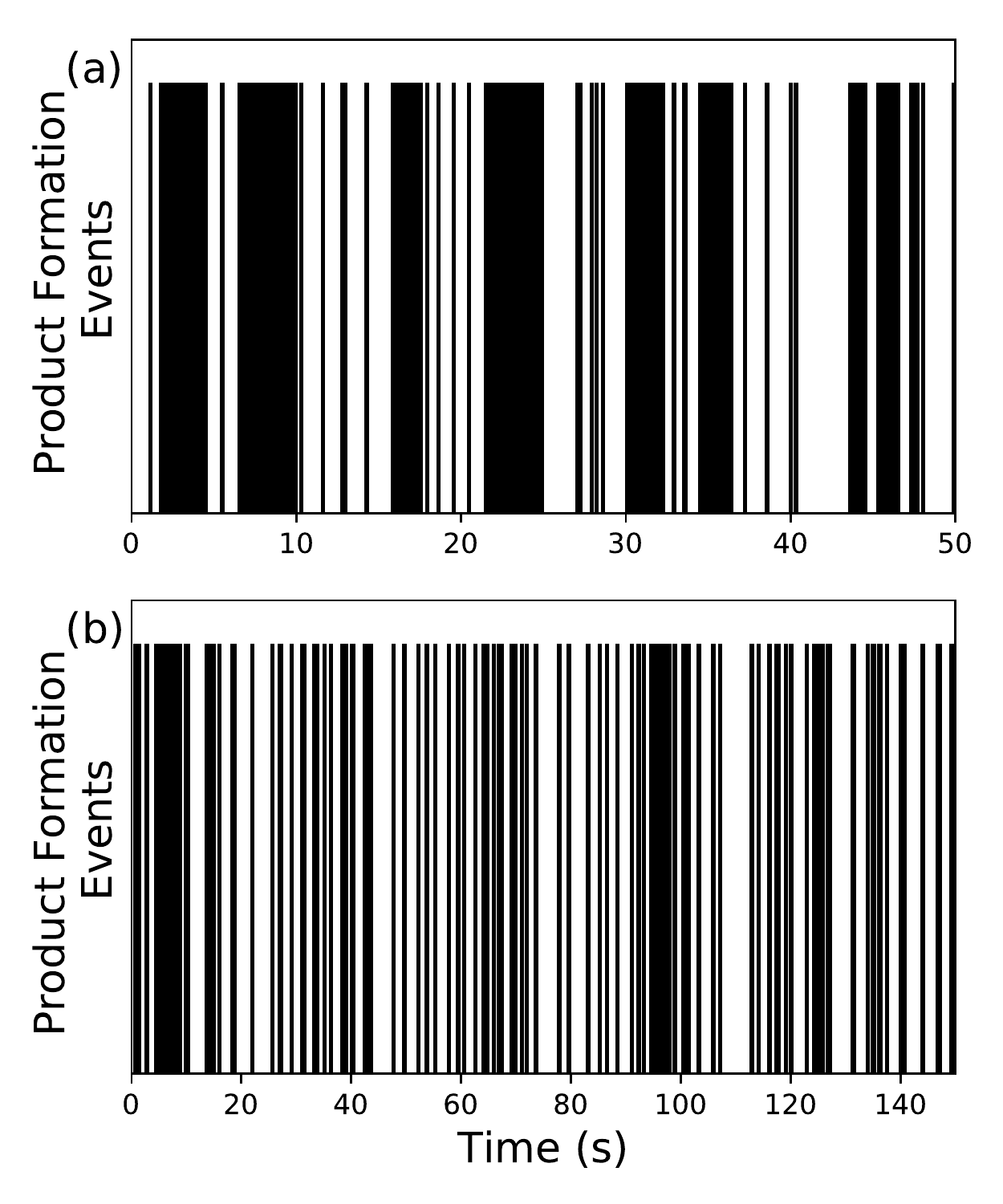}
\caption{\label{fig:allost_burst} \textbf{Time series of product arrivals stochastically simulated using Gillespie's algorithm from coupled chemical reactions shown in Fig.~\ref{fig:model} with an allosterically regulated  (a) and unregulated (b) protein}. The product arrivals are represented by vertical lines. The model with allostery exhibits a ``bursty behavior". This behavior is not observed in the absence of allostery. Parameter values used are: $k_{X+}=50s^{-1}, k_{X-}=25s^{-1}, d_{p}=0.9s^{-1}, k_{A+}=50s^{-1}, k_{A-}=25s^{-1}, k_{X^{*}+}=75s^{-1}, k_{X^{*}-}=25s^{-1}, r_{p}=50s^{-1}, s_{p}=70s^{-1} , h_{p}=50s^{-1}$ for (a) and all the same except $h_{p}=0s^{-1}$ for (b). For both the simulations we have an enzyme present in an excess of substrate (50 molecules).}
\end{figure}

For illustrative purposes, the reactions described in the processes a) to d) in Fig.~\ref{fig:model} can be simulated in a straightforward fashion using the Gillespie algorithm~\cite{gillespie_exact_1977}. Here, we simulate the product arrival events described in  Fig.~\ref{fig:model} with the help of a stochastic simulation for a sender enzyme in a system with a fixed large number of substrates. In the case of allosterically up-regulated sender enzyme (i.e., when the rate constant/s $k_{X^{{*}+}}$ or/and $d_p$ is/are greater than rate constant/s $k_{X^{{*}+}}$ or/and $r_p$), we observed ``bursts" of higher rates of product formation events in the midst of a lower rate of product formation, Fig.~\ref{fig:allost_burst}. On the other hand, without any allosteric regulation, the waiting times between product arrivals are approximately exponentially distributed (as expected). Put differently, the ratio of the mean squared to the variance in the waiting times between product arrival events which is $\approx 1.00$, for the choice of parameters specified in the caption of Fig.~\ref{fig:allost_burst} as expected for exponentially distributed waiting times. This ratio strongly deviates from unity ($\approx 17.40$) in the presence of up-allosteric regulation in the sender enzyme, Fig.~\ref{fig:allost_burst}. These stochastic simulations are also able to distinguish the variation of product formation rates of sender enzymes with the amount of substrate present in the presence of allosteric regulation, and its absence (see Supplementary-Figure.2). Analogous results are observed in the presence of down-regulation in sender enzymes. In addition, our stochastic simulations also suggests that the model is able to replicate a ``bursty" product formation in the presence of both K- and V- type allosteric regulations (see Supplementary-Figure.3).

The results of stochastic simulation suggest that in the presence of allosteric up/down regulation (usually referred as positive/negative cooperativity), the time of arrival of products exhibit a more complicated behavior as the catalytic site can exist in more than one type of state.  Next, the products generated in bursts may bind to the receiver protein downstream transmitting the signal. Due to the bursty product formation, the amount of information encoded in the up-signal, propagated by the binding events between the products and the receiver protein, is a complicated function that depends on the time-dependent interaction between the allosteric regulator site and the catalytic site of the sender enzyme.

In order to study the information encoded in the bursts of products in allosterically regulated sender enzymes, the model in Fig.~\ref{fig:model} can be further simplified under the condition of steady state. At steady state, i.e., when the amount of bound complexes of the catalytic site and allosteric regulator become time independent, the system of chemical reactions can be simplified to the follow qing:
\begin{eqnarray}
   \ce{& R + S ->[k_R] P + R} \label{red_rec_1} \\
   \ce{& T + S ->[k_T] P + T}. \label{red_rec_2}
\end{eqnarray}
Here, $R$ denotes the biochemical state of the enzyme where the functional site is in the original unbound condition ($X$) and $T$ denotes the alternate biochemical state of the sender enzyme where the catalytic site exists in its unbound condition as $X^{*}$. The rate constants for the reactions, i.e., $k_{R}=\frac{k_{X+} d_p}{k_{X-}}$ and $k_{T}=\frac{k_{X^{*}+} r_p}{k_{X^{*}-}}$ are the effective product formation rate constants for the catalytic site in states $R$ and $T$ respectively (see section 1.2 in Supplementary Information for more details).
 
 Such systems with allosterically regulated sender enzymes can be represented by two state systems--state 1 (represented by $R$ in reaction~\ref{red_rec_1}) where the catalytic site is in its original unbound condition $X$, and state 2 (represented by $T$ in reaction~\ref{red_rec_2}) where the catalytic site is in the alternate (i.e., allosterically regulated) unbound state as $X^{*}$. Moreover, as shown in Fig.~\ref{fig:model}, the switching between the states of the sender enzyme is facilitated by processes b) and d). According to these, the rate of transition of the enzyme from state $R$ to state $T$ would be proportional to $h_p$ and the rate of transition from state $T$ to $R$ would be proportional to $s_p$ (see section 1.2 for details in Supplementary Information). We note that, the state of the sender enzyme ($R$ or $T$) depends only on the state of the catalytic site (i.e., whether it is $X$ or $X^{*}$). The allosteric regulator site can be either unbound or bound with the substrate while enzyme is in any of the state.

\begin{figure}
    \centering
    \includegraphics[width=3.5in]{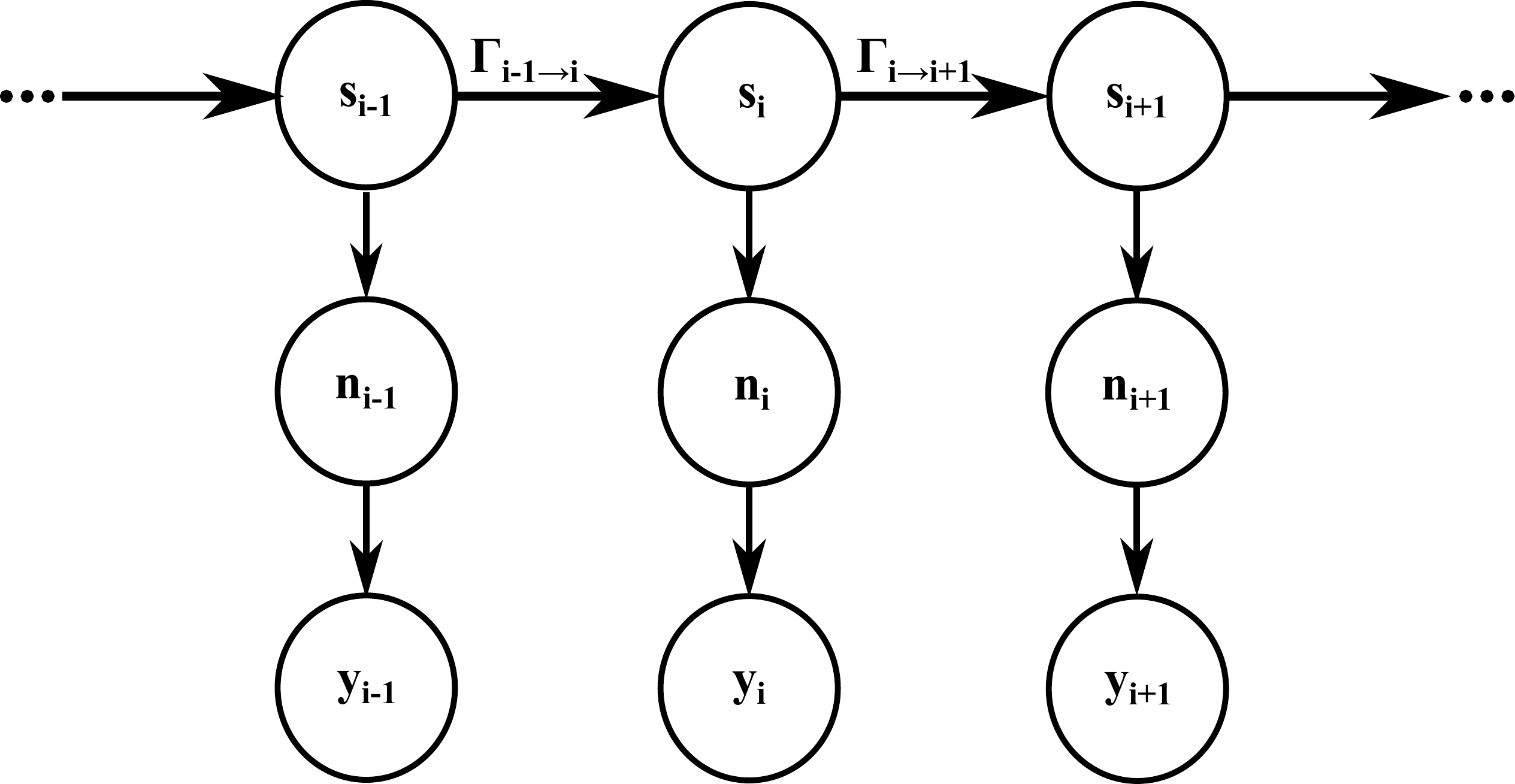}
    \caption{\label{fig:HMM} \textbf{Graphical Model describing a two state system of a sender enzyme with Poissonian emission and a receiver protein}. 
    Here the observable, $n_{i}$, is the number of products produced by a catalytic site at time level $i$ in the hidden state $s_i$. $\Gamma_{i-1\rightarrow i}$ represents the transition matrix of the states between time steps $i-1$ to $i$ and $y_i$ shows the status of the receiver protein at time interval $i$.}
\end{figure}

This model can be further simplified with the help of HMMs. In the language of HMMs, the state of the sender enzyme can be represented by a variable ($s$). Further on, in order to represent the state of the sender enzyme as a Markov chain, we discretize time into intervals of $\delta t$. The time interval $\delta t$ is selected to be small enough such that only one or no product can be produced in any time interval regardless of the state of the sender enzyme. This choice of small $\delta t$ also ensures that the sender enzyme retains its state during $\delta t$. This approximation is also backed by several studies which suggest that allostery manifests itself on timescales varying from the order of 10 ps to several nanoseconds~\cite{MCDONALD20132372}. As a result, time scales pertaining to the rates of transitions between the states of sender enzyme (i.e., $R$ to $T$ and $T$ to $R$) are several orders slower than the time scales involved in the formation of products in a given state (see Fig.~\ref{fig:HMM}). 

Therefore, the probability distribution of the states of the sender enzyme during time interval $i$ ($P(s_i)$) can be expressed as a function of the probability distribution of states during the time interval $i-1$ and the dynamics of the catalytic site and the allosteric site in the sender enzyme as~\cite{kampen_stochastic_1992}:
 \begin{eqnarray}
   P(s_{i}|s_{i-1})=\Gamma_{i-1\rightarrow i}
   \label{eq1}
\end{eqnarray}
where, $\Gamma_{i-1\rightarrow i}$ represents the transition matrix whose elements describe the probability of transition to a state $s_i$ at time interval $i$ given the state of the sender enzyme at time interval $i-1$. Moreover, once the enzyme reaches a steady state of activity, the dynamics of the above described process becomes independent in time. Therefore, under such conditions, $\Gamma_{i-1\rightarrow i}$ becomes independent of $i$ and will be referred as $\Gamma$ for the rest of the study.

The HMM model in Fig.~\ref{fig:HMM} describes the dynamics of a catalytic site existing in state $s_i$ at time step $i$ and produces products $n_i$ as described by the Poisson process given below:
\begin{eqnarray}
   P(n_i|s_i)=\frac{e^{-\lambda_{s_i}}{\lambda_{s_i}}^{n_i}}{{n_i}!}
\end{eqnarray}
where, $\lambda_{s_i}$ is the average number of products generated by the sender enzyme in state $s_i$ at time interval $i$. This parameter is closely related to the rate constants for chemical processes described in Fig.~\ref{fig:model} (see section 1.2 in Supplementary Information for more details):
\begin{eqnarray}
    \lambda_{s_i}=\begin{cases}
    d_{p}[RS]_{eq}\delta t, & \text{if $s_{i}=1$}\\
    r_{p}[TS]_{eq}\delta t, & \text{if $s_{i}=2$}.
  \end{cases}
\end{eqnarray}
Here, $[RS]_{eq}$ and $[TS]_{eq}$ are the number of bound complexes between the substrate and the catalytic site in state 1 and 2 respectively at the steady state. It should be noted that here we have used number of bound complexes as opposed to the traditional use of concentration. This change is also reflected in the units of the rate constants. Therefore, using this HMM framework, the probability of generating $n_i$ products during time interval $i$ can be written as: 
\begin{equation}
    P(n_i)=\displaystyle\sum_{s_{i}=1}^{2}P(n_i|s_i)P(s_i).
\end{equation}
Finally, we must also describe how $n_{i}$ changes the state of the receiver protein. As we select a time interval for the event to occur, at most one product can be produced in any time interval by the catalytic site. Consistent with assumptions inherent to the Gillespie simulation, we assume that diffusion occurs on timescales vastly exceeding the rate of any chemical reaction (including the product production rate). The product can then either bind to the receiver protein or disappear, by diffusing to a sink (such as an off-pathway, receiver) but it does not accumulate. Hence, the probability of the state of the receiver protein downstream in the signaling pathway (i.e., whether it is in complex with the product at time interval $i$ or not) can be given as:
 \begin{eqnarray}
    P(y_i=1|n_i)=\begin{cases}
    0, & \text{if $n_{i}=0$}\\
    p_Y, & \text{if $n_{i}=1$}
  \end{cases}
\end{eqnarray}
or,
\begin{eqnarray}
    P(y_i=0|n_i)=\begin{cases}
    1, & \text{if $n_{i}=0$}\\
    1-p_Y, & \text{if $n_{i}=1$}
  \end{cases}
\end{eqnarray}
where, $y_i$ reflects a binding event at the receiver protein during time interval $i$ such that if $y_i=1$ for a successful binding event and $y_i=0$ otherwise. Here, $p_Y$ represents the probability of a product binding at the receiver protein (see section 1.2 in  Supplementary Information for more details). Therefore, at time interval $i$, the products generated by the sender enzyme which acts as a sender of a message are received by another receiver protein downstream in the signaling pathway which encodes the information onto the $y_i$.

\section{Results and Discussion}
In the section above, we reduced the dynamical model of allosteric regulation of a sender enzyme to a simplified two-state model which can be represented by an HMM in discrete time. The model consists of a sender enzyme  with a catalytic site and an allosteric regulator site. The sender enzyme binds with substrate which leads to the generation of products. The sender enzyme can exist in two different states as $R$ (when the catalytic site exists as $X$) and $T$ (when the catalytic site exists as $X^*$). The two states differ in their dynamics of substrate binding and the rate of product formation by the catalytic site. 

The sender enzyme communicates with a receiver protein downstream in the signaling pathway with the help of products generated in time. The binding of the product with the receiver protein represents a successful reception of the signal. Finally under the condition of steady state, we can express the dynamics of switching of states of the sender protein as a function of allosteric interaction between the catalytic site and the allosteric regulator.

Now, exploiting tools from information theory, we define the amount of information conveyed by the sender enzyme to the receiver protein downstream at time interval $i$, as the mutual information (MI)~\cite{Cover:2006:EIT:1146355} between them.
\begin{equation}
    MI=\displaystyle\sum_{s_i=1}^{2}\displaystyle\sum_{y_i=0}^{1}P(s_i,y_i)\log{\frac{P(s_i,y_i)}{P(s_i)P(y_i)}} \label{MI_eq}
\end{equation}
where, $P(s_i,y_i)$ represents the joint probability distribution between $s_{i}$ and $y_{i}$ at time interval $i$. The MI can be computed for any $i$ (see section 2 for more details in Supplementary Information). Here we choose a large value of $i$, $i=500$, to ensure that the system has reached steady state dynamics and that the MI no longer varies with $i$.

In the model, the modulation of the information conveyed through product arrivals can be achieved by two different mechanisms: (i) by only changing the average number of products generated by the sender enzyme in the state $R$ or  $T$, and (ii) by altering the switching probabilities between these two states.

\begin{figure}
    \includegraphics[width=3.4in]{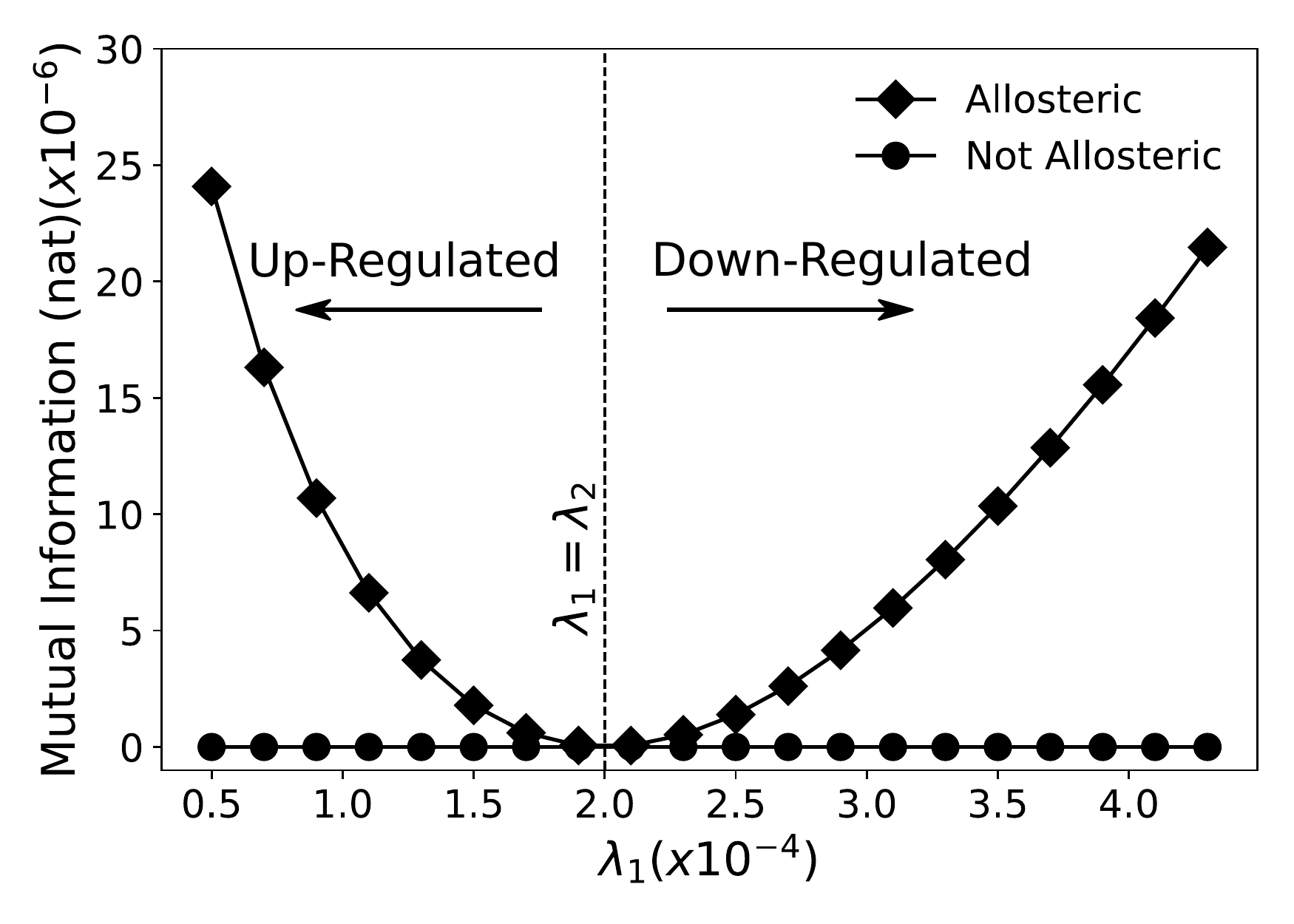}
    \caption{\label{fig:Info_lamda1} \textbf{Effect of allostery, as captured by $\lambda_1$, on the information communicated by the sender enzyme.} With allosteric up-regulation, i.e., when $\lambda_{1}<\lambda_{2}$, the time series of product arrivals encodes a larger amount of information. The dotted line represents the value of  $\lambda_{1}$ for which $\lambda_{1}=\lambda_{2}$ and where the benefit of allostery on the MI vanishes. With allosteric down-regulation, i.e., when $\lambda_{1}>\lambda_{2}$ the sender enzyme still communicates more information than in the absence of allostery. For the case of the allosteric sender enzyme, we used $\lambda_{2}=2\times10^{-4}$, and the transition probability of the sender enzyme from state $R$ to $T$, $P_{1\rightarrow 2}$ and vice versa, $P_{2\rightarrow 1}$ as 0.2 and for the case of non-allosteric sender enzyme, $P_{1\rightarrow 2}=0$. }
\end{figure}

We first investigated the effect of varying the product formation rate of the first state $R$, $\lambda_{1}$, while keeping the product formation rate of the second state $T$ and  the switching probabilities between states constant, Fig.~\ref{fig:Info_lamda1}. First, as a sanity check, we observe that without any allosteric regulation, i.e., when the second $T$ state is never visited by the sender enzyme (setting the transition probability of the sender enzyme from state $R$ to $T$ ($P_{1\rightarrow 2}$) as zero), the MI between the state of the sender enzyme and the binding events at the receiver protein is zero. This is an expected result as, in this case, the sender protein is restricted to a single state, and alternate states can no longer influence the binding of product at the receiver protein.  Therefore, no information is conveyed by the products from the sender enzyme to the receiver protein, regardless of the rate of product formation by the sender. 

Next, when $\lambda_{1}<\lambda_{2}$, i.e., in the case of up-regulation, the MI between the sender enzyme and the receiver protein is higher in contrast to the case when allosteric regulation is absent (i.e., when $P_{1\rightarrow2}=0$). However, as $\lambda_{1}$ increases and comes closer to the value of $\lambda_{2}$, the difference between the product formation rates in the two states of the sender enzyme decreases and so does the MI until it drops down to zero when $\lambda_{1} = \lambda_{2}$. This observation is counter-intuitive as it suggests that the MI between the sender enzyme and the receiver protein decreases with an increase in the average rate of product formation. Equally counter-intuitively, the information communicated by the sender to the receiver rises once again when $\lambda_{1}>\lambda_{2}$, i.e., for the case of down-regulation.

\begin{figure}[h!]
    \includegraphics[width=3.4in]{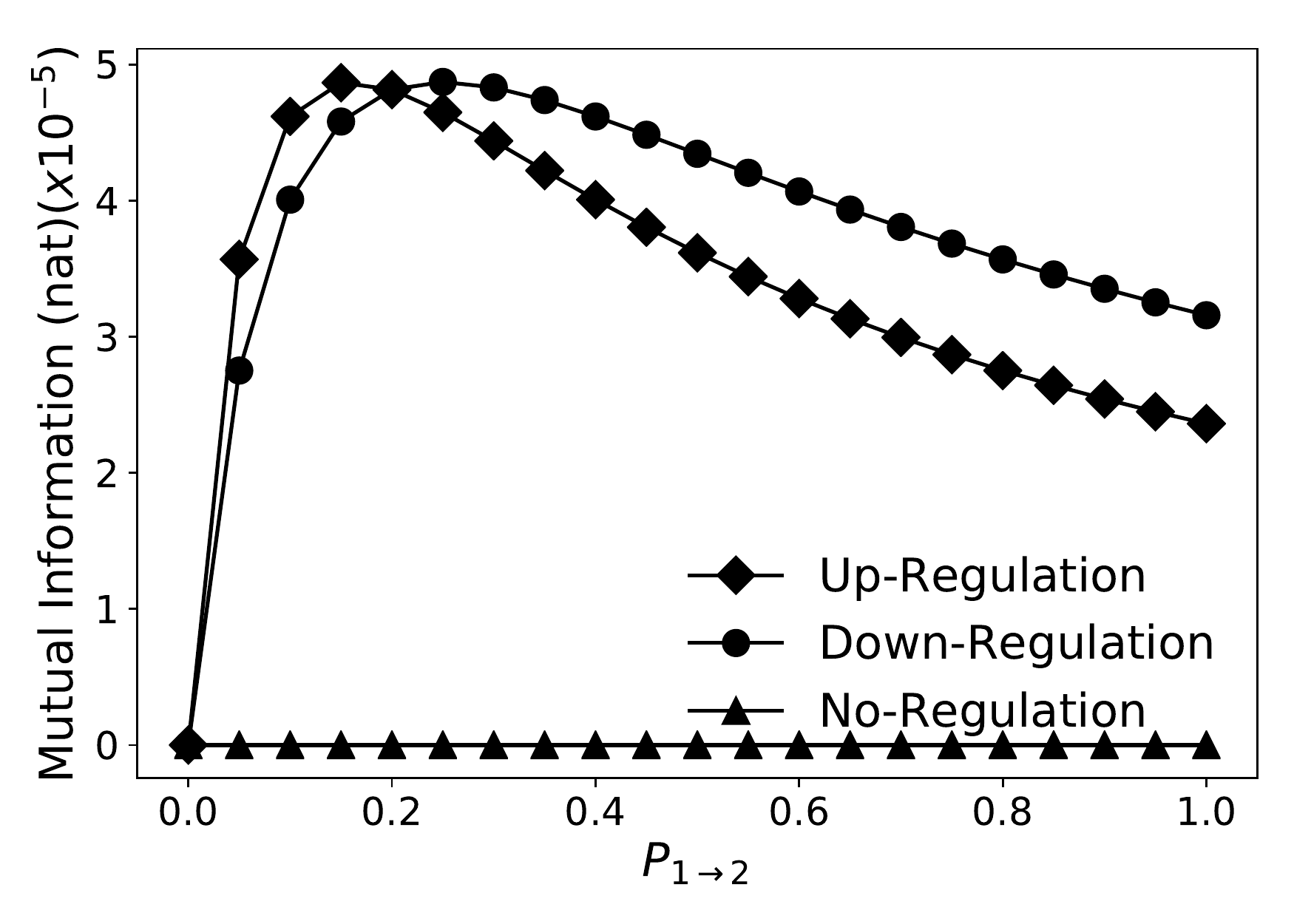}
    \caption{\label{fig:Info_P12} \textbf{Effect of changing the switching probability from state $R$ to $T$, $P_{1\rightarrow2}$ on the information communicated by the sender enzyme for different types of allosteric regulations}. We observe that without any allosteric regulation, the information encoded in the product time series do not vary with switching probabilities. However, with up-regulation, i.e., when $\lambda_{1}(1\times10^{-4})<\lambda_{2}(4\times10^{-4})$, the sender enzyme communicates more information as the probability of it switching to an up-regulated state increases. Counter-intuitively, a down-regulated sender enzyme, i.e., when $\lambda_{1}(4\times10^{-4})>\lambda_{2}(1\times10^{-4})$, is still able to communicate a larger amount of information for a wide range of switching probabilities. For all plots, we fix the transition probability from state $T$ to $R$, $P_{2\rightarrow1}$ to $0.08$.}
\end{figure}

Second, we analyzed the variation of the information transmitted with allosteric regulation by varying the probability of switching of the state of the sender enzyme from $R$ to $T$, i.e., $P_{1\rightarrow 2}$, while holding other parameters (i.e., $\lambda_{1}$, $\lambda_{2}$, and $P_{2\rightarrow1}$) fixed, Fig.~\ref{fig:Info_P12}. As a sanity check, we first observed the amount of information conveyed in the absence of allosteric regulation, i.e., when $\lambda_{1} = \lambda_{2}$. As expected (similar to the case observed above), the signal no longer carries any MI when the effect of allostery is absent, and is also insensitive to the value of $P_{1\rightarrow 2}$. Which implies that the binding events at the receiver protein would be independent of the state the sender enzyme and, as a result, there is no MI between them. On the other hand, for up- and down-regulated proteins, the sender enzyme conveys minimum information while being restricted to only one state (i.e., when $P_{1\rightarrow 2}=0$). However, as $P_{1\rightarrow 2}$ grows, with up-regulation, the sender enzyme is able to access a state with higher average production and therefore conveys a larger amount of information to the receiver protein. Although, unexpectedly, with down-regulation, the sender enzyme is still able to transmit a larger amount of information for a range of switching probabilities. This counter-intuitive observation illustrates once more that information transmitted is not a mere function of the amount of product available to the receiver protein but also depends on the switching kinetics between the allosteric states of the sender enzyme.

\section{Conclusion}
Here we investigated a minimal model of allostery and quantified the information communicated by the catalytic site of an allosterically regulated sender enzyme to another receiver protein further down in the signaling pathway through time-varying product formation rates. Our choice of a two-state model is a matter of convenience; our formalism readily generalizes to more complex kinetic models.

More importantly, we found that an allosterically regulated enzymes may convey a larger amount of information as compared to an enzyme with no allosteric regulation by decreasing its net rate of product formation.

This suggests that allostery may provide means to control the information encoded in the time of arrival of products in a way that goes beyond the energetically demanding ``more product, better signal" exploitative paradigm. That is, allostery may provide ``lower signal but more information". The possibility of parameter fine-tuning to communicate more information is especially relevant given allostery's key role in protein evolution~\cite{modi_tushar_ancient_2018,nussinov_principles_2014,townsend_role_2015}. It opens the possibility that nature may fine-tune allosteric parameters (including switching rates between states as well as production rates) to adapt/evolve its signaling pathways in response to external stimuli warranting exploratory (high information/low signal) or exploitative (low information/high signal) strategies~\cite{hills_exploration_2015}.

\section{Acknowledgements}
This project is supported by ARO grant W911NF-17-1-0162  
on ``Multi-Dimensional and Dissipative Dynamical Systems: Maximum Entropy as a Principle for Modeling Dynamics and Emergent Phenomena in Complex Systems". SBO and TM also acknowledge  the NSF-MCB Award1715591 and Moore Foundation. We also thank Drs. Ioannis Sgouralis and Sean Seyler for helpful discussions.

\newpage
\bibliography{Allosteric_enz}

\end{document}


\preprint{PRL}
\title{Information propagation in time through allosteric signaling -- Supplementary Information}

\author{Tushar Modi}
\affiliation{Department of Physics, Arizona State University, Tempe, Arizona 85287, USA}

\author{S. Banu Ozkan}
\affiliation{Department of Physics, Arizona State University, Tempe, Arizona 85287, USA}
\affiliation{co-corresponding}

\author{Steve Press\'e}
\affiliation{Department of Physics, Arizona State University, Tempe, Arizona 85287, USA}
\affiliation{School of Molecular Sciences, Arizona State University, Tempe, Arizona 85287, USA}
\affiliation{co-corresponding}

\date{\today}

\maketitle
\section{Model for allosteric regulation}
\subsection{Coupled chemical reactions for an alloterically regulated sender enzyme}
\begin{figure}[h!]
    \centering
    \includegraphics[width=3.5in]{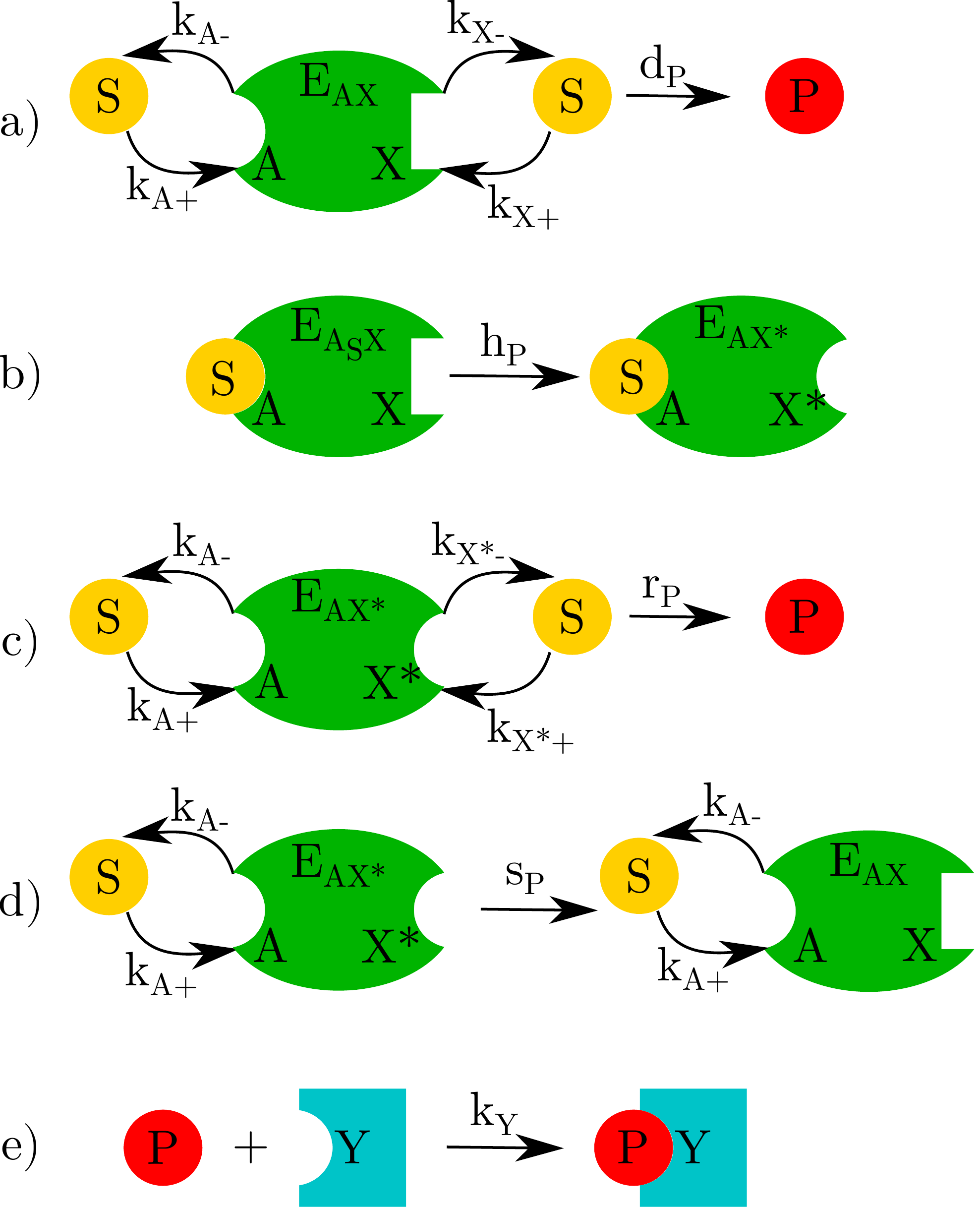}
    \caption{\label{fig:model} \textbf{Schematic diagram of the model of an allosterically regulated enzyme.} It shows a sender enzyme $E_{AX}$ with catalytic site $X$ interacting with substrates $S$ to make product $P$. The enzyme also contains an allosteric regulator site $A$ which can interact with the substrates to change the state of the catalytic site to $X^{*}$ which can also interact with the substrate to make products $P$. The products can then bind to the receiver protein $Y$ downstream in the signaling pathway.}
\end{figure}

In the main text, we proposed a general model for an enzyme $E_{AX}$ which describes the interaction of a catalytic site $X$--binding with a substrate $S$ to form product $P$--whose binding and catalytic activity is modulated by another binding site $A$ through allosteric regulation, Supplementary Figure.~\ref{fig:model}.

\begin{figure}
\includegraphics[width=3.4in]{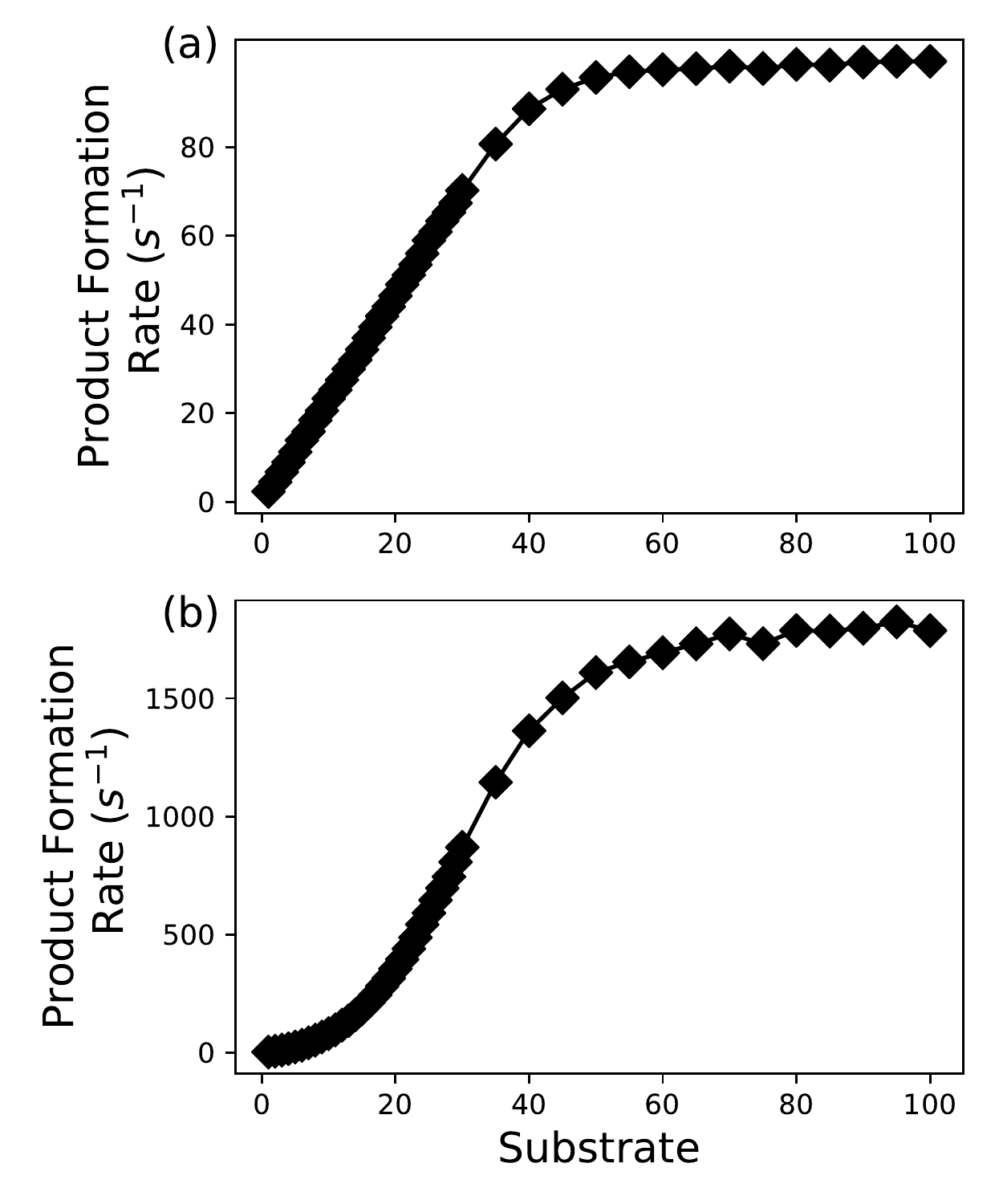}
\caption{\label{fig:prod_rates} \textbf{Variation in rates of product formation with substrate concentration}. We observe that the model without allosteric regulation, (a) recapitulates the dynamics commonly observed in proteins following Michaelis-Menton Kinetics. On the other hand, when allosteric regulation is present, (b) the rate of product formation is non-linear at lower concentrations of substrate, a signature of cooperativity due to allosteric regulation. The following parameter values were used: $k_{X+}=50s^{-1}, k_{X-}=25s^{-1}, d_{p}=5s^{-1}, k_{A+}=50s^{-1}, k_{A-}=25s^{-1}, k_{X^{*}+}=75s^{-1}, k_{X^{*}-}=50s^{-1}, r_{p}=10000s^{-1}, s_{p}=50s^{-1}, h_{p}=0s^{-1}$ for (a) with a change of  $h_{p}$ to $h_{p}=100s^{-1}$ for (b). The system was simulated for 30s and rates for each substrate concentration were averaged over 1000 independent runs. For both the simulations we have an enzyme present in an excess of substrate (50 molecules).}
\end{figure}

The generation of products through this set of coupled chemical reactions are realized using Gillespie's algorithm~\cite{gillespie_exact_1977}. Firstly, as a proof of concept, we analyze the rate of product formation in our model for the sender enzyme with no allosteric regulation between catalytic sites $X$ and $A$, i.e., when $h_{p}=0s^{-1}$, Supplementary Figure.~\ref{fig:prod_rates}a. We observe that, as expected, without any allosteric regulation, the system of chemical reactions reduce to  Michaelis-Menton kinetics~\cite{johnson_original_2011} such that the rate of product formation increases linearly at lower substrate concentrations and then saturates to a constant value at higher substrate concentration. On the other hand, when allosteric regulation is present, i.e., $h_{p}>0s^{-1}$, the production rate exhibits a non-linear relationship at lower substrate concentration, commonly observed in enzymes with allosteric regulation~\cite{monod_nature_1965,koshland_comparison_1966,cooper_allostery_1984}, Supplementary Figure.~\ref{fig:prod_rates}b.

\begin{figure}
\includegraphics[width=3.4in]{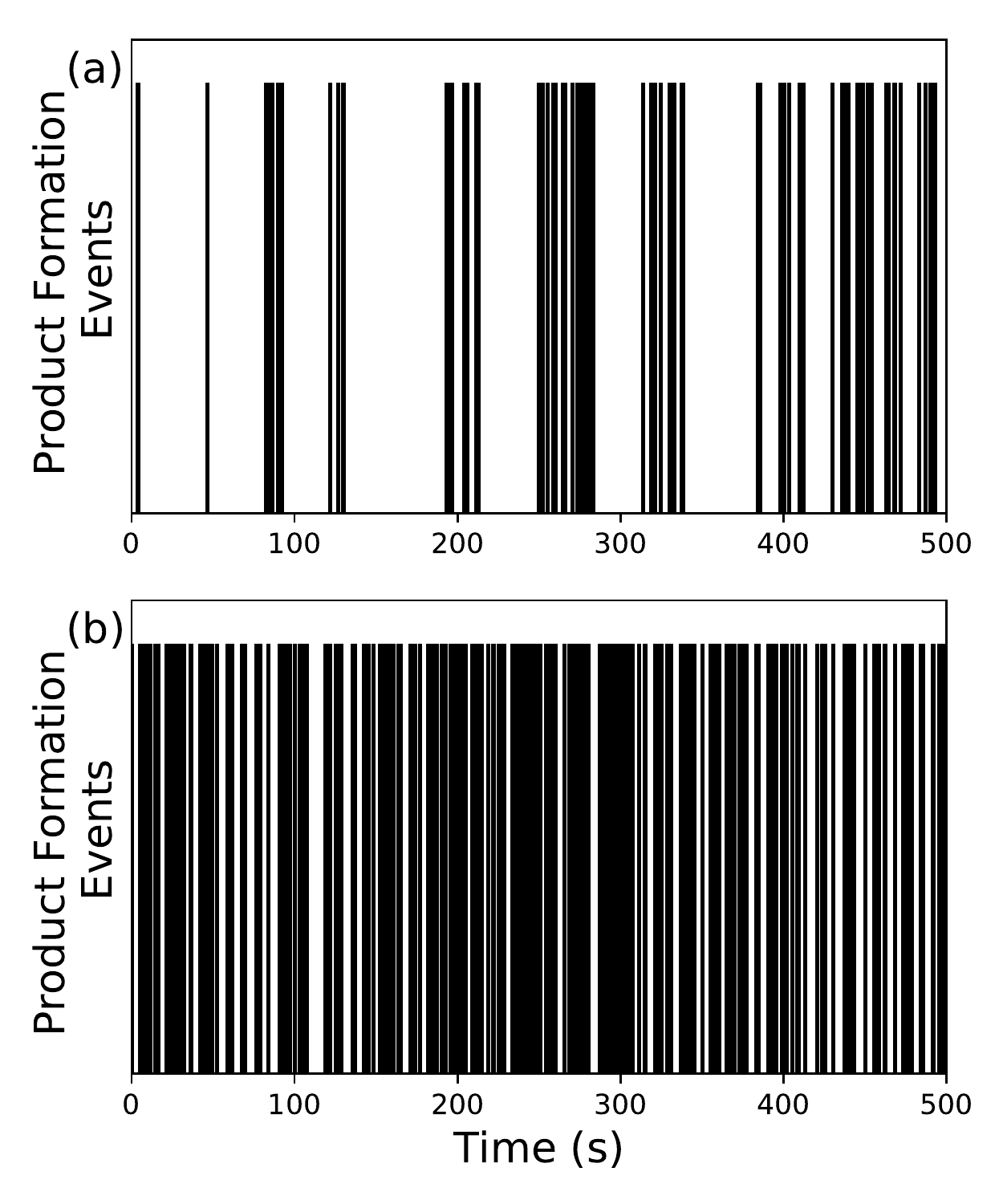}
\caption{\label{fig:k_type} \textbf{Bursts of products observed in K-type allostery}. We observe that the model with K-type allosteric regulation, (a) shows a ``bursty" formation of products which is not observed in the model without allosteric regulation, (b). Vertical lines represent product formation events. The following parameter values were used: $k_{X+}=10000s^{-1}, k_{X-}=10s^{-1}, d_{p}=0.5s^{-1}, k_{A+}=1000s^{-1}, k_{A-}=25s^{-1}, k_{X^{*}+}=0.01s^{-1}, k_{X^{*}-}=1000s^{-1}, r_{p}=0.5s^{-1}, s_{p}=0.04s^{-1}, h_{p}=5000s^{-1}$ for (a) and $h_{p}=0s^{-1}$ for (b). For both the simulations we have an enzyme present in an excess of substrate (50 molecules).}
\end{figure}

\begin{figure}
\includegraphics[width=3.4in]{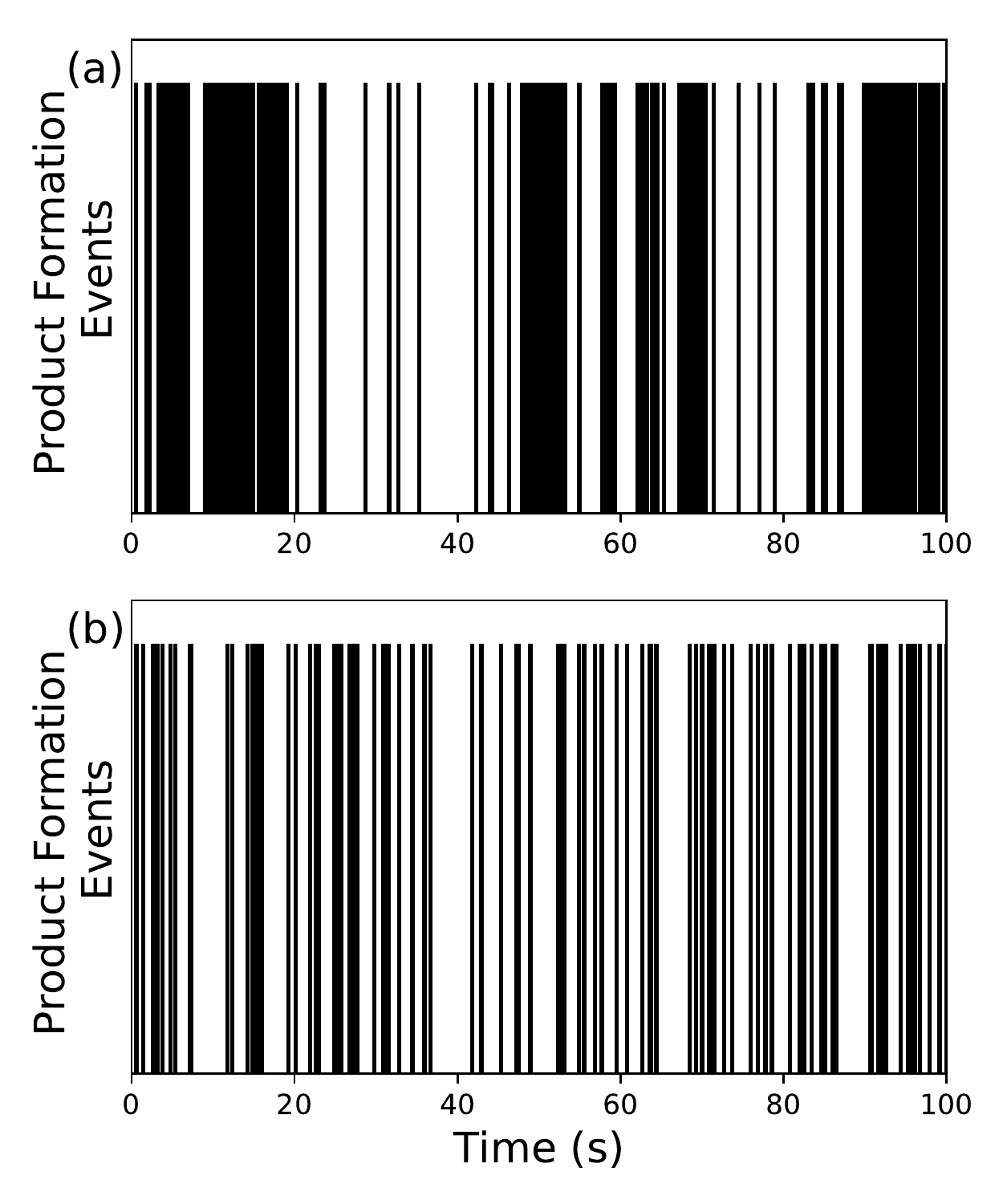}
\caption{\label{fig:v_type} \textbf{Bursts of products observed in V-type allostery}. We observe that the model with V-type allosteric regulation, (a) shows a ``bursty" formation of products which is not observed in the model without allosteric regulation, (b). Vertical lines represent product formation events. The following values of the parameters were used: $k_{X+}=50s^{-1}, k_{X-}=25s^{-1}, d_{p}=0.3s^{-1}, k_{A+}50s^{-1}, k_{A-}=25s^{-1}, k_{X^{*}+}=50s^{-1}, k_{X^{*}-}=25s^{-1}, r_{p}=10s^{-1}, s_{p}=70s^{-1} and , h_{p}=50s^{-1}$ for (a) and $h_{p}=0s^{-1}$ for (b). For both the simulations we have an enzyme present in an excess of substrate (50 molecules).}
\end{figure}

As discussed in the main text, the model proposed in Supplementary Figure.~\ref{fig:model} above produces a unique ``bursty" emission of products when allosteric regulation is present. The model also captures this ``bursty" production arising from different microscopic models of allosteric regulation~\cite{traut_enzyme_2014}, i.e., namely K-type models (when allosteric regulation manifests itself by modulating the catalytic site's binding affinity with the substrate, Supplementary Figure.~\ref{fig:k_type}) and V-type models (when allosteric regulation affects the rate by which the bound complex of catalytic site reduces to give products, Supplementary Figure.~\ref{fig:v_type}).

The products generated by the enzyme, then interact with a receiver protein ($Y$) downstream in the signaling pathway. In this model, a successful binding event signifies transmission of a signal by the enzyme down the signaling pathway:
\begin{eqnarray}
   \ce{ P + Y ->[k_Y] PY}.
\end{eqnarray}

\subsection{Coupled chemical reactions for allosteric regulation reduce to a two state Hidden Markov model}
As the system reaches steady state, i.e., when the forward and backward reaction of complex formation between substrate and the sender enzyme are equal, the concentrations of bound complexes do not change with time. Therefore, for reactions in step (a) in Supplementary Figure.~\ref{fig:model}, when the rate of formation of $[E_{AX_{S}}]$ (and $[E_{A_{S}X_{S}}]$) and its rate of degradation are equal we can write:
\begin{eqnarray}
    \frac{[E_{AX_{S}}]_{eq}}{[E_{AX}]_{eq}[S]_{eq}}&=&\frac{k_{X+}}{k_{X-}} \nonumber\\
    \frac{[E_{A_{S}X_{S}}]_{eq}}{[E_{A_{S}X}]_{eq}[S]_{eq}}&=&\frac{k_{X+}}{k_{X-}}
\end{eqnarray}
where, $E_{AX_{S}}$ represents the sender enzyme $E$ with the unbound allosteric site $A$ and the catalytic site $X$ bound to a substrate, and the subscript $eq$ stands for the concentration at the steady state. Similarly, when the rate of formation of bound allosteric site, $[E_{A_{S}X}]$ (and $[E_{A_{S}X_{S}}]$) and its rate of degradation are equal, we can write:
\begin{eqnarray}
    \frac{[E_{A_{S}X}]_{eq}}{[E_{AX}]_{eq}[S]_{eq}}&=&\frac{k_{A+}}{k_{A-}} \nonumber\\
    \frac{[E_{A_{S}X_{S}}]_{eq}}{[E_{AX_{S}}]_{eq}[S]_{eq}}&=&\frac{k_{A+}}{k_{A-}}. \label{eq:rate_cond1}
\end{eqnarray}
Similarly, when then rate of formation of complex between the catalytic site in its alternate state ($X^{*}$) and the substrate, i.e. $[E_{AX^{*}_{S}}]$ (and $[E_{A_{S}X^{*}_{S}}]$), and its rate of degradation are equal, we can write:
\begin{eqnarray}
    \frac{[E_{AX^{*}_{S}}]_{eq}}{[E_{AX^{*}}]_{eq}[S]_{eq}}&=&\frac{k_{X^{*}+}}{k_{X^{*}-}} \nonumber\\
    \frac{[E_{A_{S}X^{*}_{S}}]_{eq}}{[E_{A_{S}X^{*}}]_{eq}[S]_{eq}}&=&\frac{k_{X^{*}+}}{k_{X^{*}-}}. \label{eq:rate_cond2}
\end{eqnarray}
For simplicity and clarity, we refer to the sender enzymes with its unbound catalytic site in its original state ($X$) as $R$ and the sender enzyme with its unbound catalytic in its alternate state ($X^{*}$) as $T$. Therefore, $RS$ and $TS$ would represent their bound forms respectively.

Given the initial concentration of the sender enzyme, $[E]_{initial}$, mass balance implies an additional condition on the steady state concentrations of the chemical species as shown below:
\begin{eqnarray}
    {[E]}_{initial}&=&[E_{AX}]_{eq}+[E_{AX_{S}}]_{eq}+[E_{A_{S}X}]_{eq}+[E_{A_{S}X_{S}}]_{eq}+ \nonumber \\
    &&[E_{AX^{*}}]_{eq}+[E_{AX^{*}_{S}}]_{eq}+[E_{A_{S}X^{*}}]_{eq}+[E_{A_{S}X^{*}_{S}}]_{eq} \nonumber \\
    {[E]}_{initial}&=&[R]_{eq}+[RS]_{eq}+[T]_{eq}+[TS]_{eq}. \label{eq:mass_cond}
\end{eqnarray}
Combining Eq.~\ref{eq:mass_cond} with Eqs.~\ref{eq:rate_cond1} and ~\ref{eq:rate_cond2} with the condition that substrate is present in excess, i.e., $[S]_{eq} \approx [S]_{initial}$, we can reduce the model, to a two state model as shown below:
\begin{eqnarray}
    \ce{& R + S <=>[k_{X+}][k_{X-}] RS ->[d_p] P + R} \label{reac_1} \\
    \ce{& T + S <=>[k_{X^{*}+}][k_{X^{*}-}] TS ->[r_p] P + T}. \label{reac_2}
\end{eqnarray}
At steady state, for reaction~\ref{reac_1} the rate of formation of $RS$ and its degradation will be equal, i.e.:
\begin{equation}
    \frac{d[RS]_{eq}}{dt}=k_{X+}[R]_{eq}[S]_{eq}=k_{X-}[RS]_{eq}.
\end{equation}
Therefore,
\begin{equation}
    [RS]_{eq}=\frac{k_{X+}}{k_{X-}}[R]_{eq}[S]_{eq}. \label{eq:temp1}
\end{equation}
Also, in reaction~\ref{reac_1} and Eq.~\ref{eq:temp1}:
\begin{eqnarray}
    \frac{d[P]}{dt}&=&d_{P}[RS]_{eq} \\
    \frac{d[P]}{dt}&=&\frac{k_{X+}d_{P}}{k_{X-}}[R]_{eq}[S]_{eq} \\
    \frac{d[P]}{dt}&=&k_{R}[R]_{eq}[S]_{eq}
\end{eqnarray}
where, $k_R$ can be treated as the effective forward rate for the production of products in reaction~\ref{reac_1}. Following a similar logic reactions ~\ref{reac_1} and~\ref{reac_2} can be expressed in the two state system as:
\begin{eqnarray}
    \ce{& R + S ->[k_R] P + R} \label{red_model_1} \\
    \ce{& T + S ->[k_T] P + T} \label{red_model_2}
\end{eqnarray}
where, $k_{R}=\frac{k_{X+}}{k_{X-}}d_p$ and $k_{T}=\frac{k_{X^{*}+}}{k_{X^{*}-}}r_p$ are the effective product formation rate constants for the generation of products from the sender enzyme in the two states ($R$ and $T$ respectively). In addition, the sender enzyme can also switch between states as described by processes b) and d) in Supplementary Figure.~\ref{fig:model}.

In order to computationally study such two state models, we first discretize time into smaller intervals ($\delta t$). The length of the interval is chosen such that within a given interval, the sender enzyme can generate only one product at most, regardless of whatever state it is in. The reason for doing so is described in the main text in detail. Afterwards, we exploit the use of Hidden Markov Models (HMM)~\cite{rabiner_introduction_1986, seymore_learning_1999} to describe this two state system where the state of the sender enzyme at time interval $i$ can be represented as $s_i$ which can take any value between 1 (for state $R$) and 2 (for state $T$). If at some time interval $i$, the sender enzyme is in state 1, the probability of remaining in state 1, $P_{1\to1}$, is proportional to $[R]_{eq}[S]_{eq}k_R$ from Eq.~\ref{red_model_1}. Whereas, the probability for switching its state from state 1 to 2, $P_{1\to2}$ will proportional to $h_p$ time the probability of having a bound allosteric site, Supplementary Figure.~\ref{fig:model}b), i.e., $h_p\frac{[E_{AX}]_{eq}[S]_{eq}k_{A+}}{k_{A-}}$. On the other hand, if the sender enzyme is in state 2, the probability to remain in the state $P_{2\to2}$ is proportional to $[T]_{eq}[S]_{eq}k_T$ and the probability to switch, $P_{2\to1}$ is proportional to $s_p$ from Supplementary Figure.~\ref{fig:model}d). These probabilities can be explicitly calculated using appropriate normalization conditions, i.e., using $P_{1\to1}+P_{1\to2}=1$ and $P_{2\to1}+P_{2\to2}=1$. These can be used to calculate the transition probability matrix for a catalytic site at steady state as:
\begin{equation}
    \Gamma =
    \begin{bmatrix}
       P_{1\to1} & P_{1\to2}\\[0.3em]
       P_{2\to1} & P_{2\to2}\\[0.3em]
    \end{bmatrix} \label{dynamics} \nonumber \\
\end{equation}
where,
\begin{eqnarray}
     P_{1\to1}&=&\frac{[R]_{eq}[S]_{eq}k_R}{[R]_{eq}[S]_{eq}k_R+h_p\frac{[E_{AX}]_{eq}[S]_{eq}k_{A+}}{k_{A-}}} \nonumber \\
     P_{1\to2}&=&\frac{h_p\frac{[E_{AX}]_{eq}[S]_{eq}k_{A+}}{k_{A-}}}{[R]_{eq}[S]_{eq}k_R+h_p\frac{[E_{AX}]_{eq}[S]_{eq}k_{A+}}{k_{A-}}} \nonumber \\
     P_{2\to1}&=&\frac{s_p}{[T]_{eq}[S]_{eq}k_T+s_p} \nonumber \\
     P_{2\to2}&=&\frac{[T]_{eq}[S]_{eq}k_T}{[T]_{eq}[S]_{eq}k_T+s_p}.
\end{eqnarray}

Afterwards, the catalytic site in state $s_i$ can generate $n_i$ number of products in the time interval $i$. Since the time interval selected is small such that the catalytic site can produce only one product in a given time interval at most, Eq.~\ref{eq:emm_cond}.
\begin{equation}
    P(n_i > 1 |s_i) \approx 0. \label{eq:emm_cond}
\end{equation}
The product molecules generated can then interact with a receiver protein ($Y$) downstream in the signaling pathway. However, if a product fails to interact with the receiver protein $Y$, then due to our selection of a very small time interval, it is safe to assume that it diffuses away and does not accumulate. Another benefit of choosing a smaller time scale is that typical allosteric interactions occur at much slower time intervals as compared to time scales involved with product formation events. Therefore, during a small time interval, it is safe to approximate that the sender enzyme will not be changing its state but only participating in product formation events. This model can be shown schematically as in Supplementary Figure.~\ref{fig:HMM}.
\begin{figure}
    \centering
    \includegraphics[width=4in]{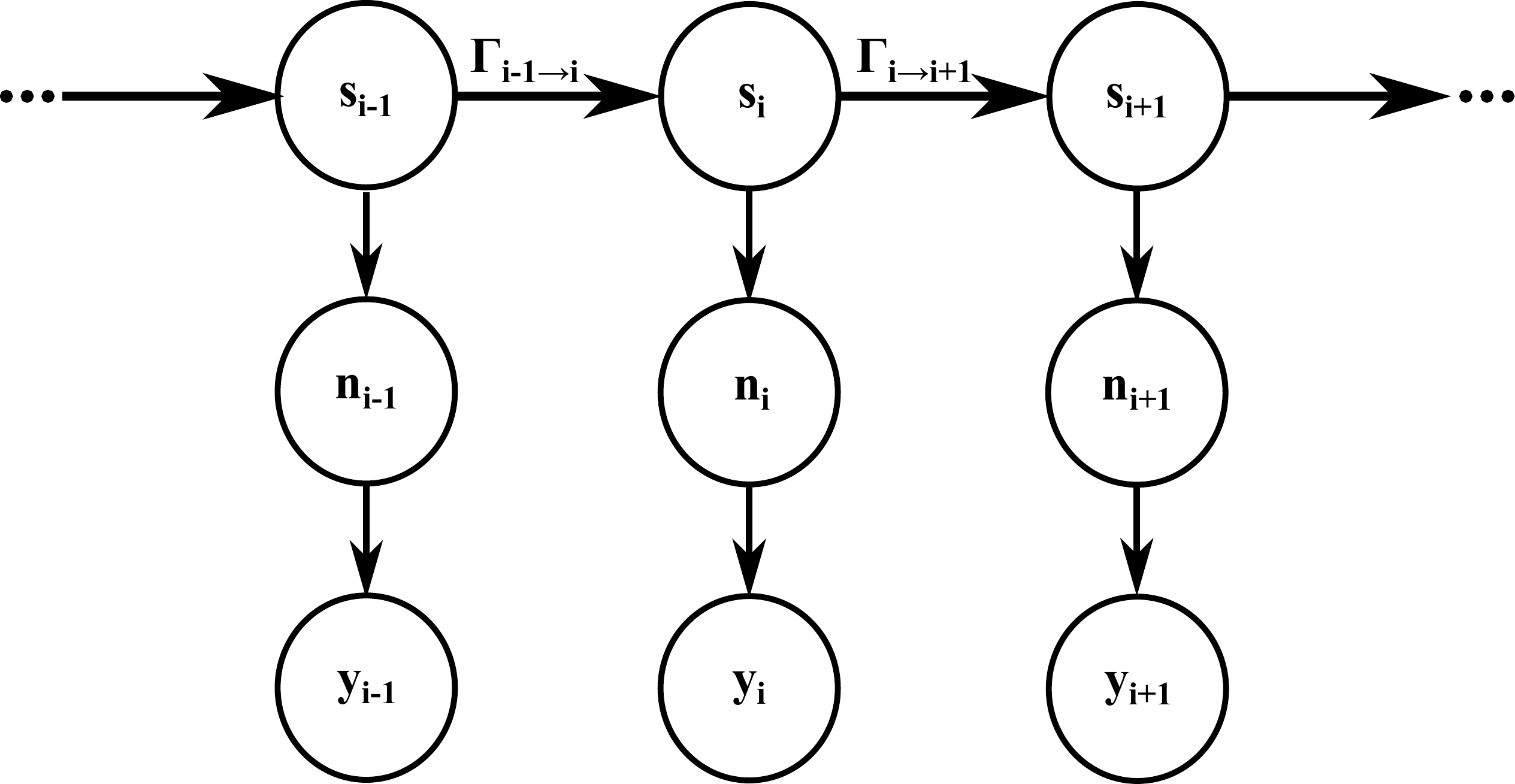}
    \caption{\label{fig:HMM} \textbf{Graphical Model describing a two state system of a sender protein with Poissonian emission and a receiver protein}.
    Here the observable, $n_{i}$, is the number of products generated by a sender enzyme at time level $i$ in the hidden state $s_i$. $\Gamma_{i-1\rightarrow i}$ represents the transition matrix of the states between time steps $i-1$ to $i$ and $y_i$ shows the status of the receiver protein at time interval $i$.}
\end{figure}

For the first layer of the HMM model, the Markov dynamics of switching from state $s_i$ at time interval `$i$' to state $s_{i+1}$ at time interval `$i+1$' can be described as:
\begin{equation}
   P(s_{i+1},s_{i})=\Gamma_{i+1 \rightarrow i} P(s_{i}).
\end{equation}
Here, $P(s_{i})$ represents the probability distribution of the state occupied by the sender enzyme at a time interval `$i$'. Moreover on account of our steady state assumptions, the elements of the transition matrix, $\Gamma$ remain constant as described in Eq.~\ref{dynamics}, therefore, $\Gamma_{i+1 \rightarrow i} = \Gamma $: $\forall i \in {1,2,...,M}$ for $M$ time steps where $M \rightarrow \infty$. 

The second layer of HMM model represents the bursts of products produced by the sender enzyme in state $s_i$ at a given time interval `$i$'. The probability distribution describing the process can expressed as:
\begin{eqnarray}
   P(n_i|s_i)=\frac{e^{-\lambda_{s_i}}{\lambda_{s_i}}^{n_i}}{{n_i}!}
\end{eqnarray}
where, $\lambda_{s_i}$ is the average number of products formed by the sender enzyme in state $s_i$. It depends on the steady state concentration of the bound complex in that state $s_i$ as well as the rate parameters as follows:
\begin{eqnarray}
    \lambda_{S_i}=\begin{cases}
    d_{p}[RS]_{eq}\delta t, & \text{if $s_{i}=1$}\\
    r_{p}[TS]_{eq}\delta t, & \text{if $s_{i}=2$}.
  \end{cases}
\end{eqnarray}
This can be used to calculate the probability of observing $n_i$ products by summing over all the possible values as $s_i$ as:
\begin{equation}
    P(n_i)=\displaystyle\sum_{s_i=1}^{2}P(n_i|s_i).
\end{equation}
Finally, the last layer of HMM model depicts the receiver process which involves the products generated in the last layer interacting with another receiver protein, $Y$, downstream in the signaling pathway. The process which describes the interaction between the two can be modeled with complexities describing the biology of the process, however, for simplicity and concreteness, we will be using a simple binomial process to describe the activation of receiver by product as:
\begin{eqnarray}
    P(y_i=1|n_i)=\begin{cases}
    0, & \text{if $n_{i}=0$}\\
    p_Y, & \text{if $n_{i}=1$}
  \end{cases}
\end{eqnarray}
and,
\begin{eqnarray}
    P(y_i=0|n_i)=\begin{cases}
    1, & \text{if $n_{i}=0$}\\
    1-p_Y, & \text{if $n_{i}=1$}
  \end{cases}
\end{eqnarray}
where, the random variable $y_i$ tells whether the receiver protein is in complex with the product or not during time interval `$i$', i.e., if the signal has successfully transmitted or not. $p_Y$ is a real number less than 1 which depicts the probability of activation of receiver in the presence of the product. Moreover, for simplicity, the dynamics of receiver are not included in the model. Hence, at the beginning of each time interval, the receiver is assumed to be in the refreshed state waiting to receive a new product. The probabilities shown above describes the dynamics of the model which will be used for calculating the amount of information encoded by the sender enzyme while communicating with the receiver protein downstream in the signaling pathway.

\section{Calculating the information transmitted by the sender enzyme to the receiver protein.}

The amount of information communicated by the sender enzyme in the probabilistic model of allosteric regulation can be calculated by the use of Mutual Information (MI)~\cite{Cover:2006:EIT:1146355}. Here we propose to calculate the mutual information between the state of the sender enzyme at some time interval `$i$' and the state of the receiver protein in the same time interval. It can be expressed as:
\begin{equation}
    MI=\displaystyle\sum_{s_i =1}^{2}\displaystyle\sum_{y_i=0}^{1}P(s_i,y_i)\log{\frac{P(s_i,y_i)}{P(s_i)P(y_i)}}. \label{eq:mi}
\end{equation}
According to the model, MI quantifies the amount of information encoded by the catalytic site dynamics in the form of arrival times of products which is transferred to a receiver protein in turn binding with it. The expression in Eq.~\ref{eq:mi}, $P(s_i)$ is the probability of the state of sender enzyme, $P(y_i)$ is the probability of the state of receiver protein and $P(s_i,y_i)$ is their joint probability. The expression is summed over all the states of the sender enzyme and the receiver protein, i.e., $s_i\in[1,2]$ and $y_i\in[0,1]$

In order to compute the expression above in Eq.~\ref{eq:mi}, the joint probability can be further reduced analytically as:
\begin{eqnarray}
    P(s_i,y_i)&=&P(y_i|s_i)P(s_i) \nonumber \\
    &=&\displaystyle\sum_{n_i=0}^{1}P(y_i,n_i|s_i)P(s_i)
\end{eqnarray}
where, due to the strategic choice of $\delta t$, $n_i\in[0,1]$. Further on,
\begin{eqnarray}
    P(s_i,y_i)&=&\displaystyle\sum_{n_i=0}^{1}P(y_i,n_i|s_i)P(s_i) \nonumber \\
    &=&\Big[\displaystyle\sum_{n_i=0}^{1}P(y_i|n_i)P(n_i|s_i)\Big]P(s_i). \label{eq:joint}
\end{eqnarray}
Here, $P(y_i|n_i)$ is the receiver process described earlier and $P(n_i|s_i)$ describes the process for the formation of products. $P(s_i)$ can be calculated as:
\begin{eqnarray}
    P(s_i)&=&\displaystyle\sum^{2}_{s_{i-1}=1}P(s_i|s_{i-1})P(s_{i-1}). \label{eq:dyn}
\end{eqnarray}
In this expression $P(s_i|s_{i-1})$ can be obtained from the switching dynamics represented by the Markov process representing the catalytic site dynamics of the sender enzyme. This equation can be computed in an iterative fashion to calculate $P(s_i)$ for some given initial conditions of the active sites, i.e., $P(s_o)$. Moreover,
\begin{eqnarray}
    P(y_i)&=&\displaystyle\sum_{n_i=0}^{1}P(y_i,n_i) \nonumber \\
    &=&\displaystyle\sum_{n_i=0}^{1}P(y_i|n_i)P(n_i) \nonumber \\
    &=&\displaystyle\sum_{n_i=0}^{1}P(y_i|n_i)\Big[\displaystyle\sum_{s_i=1}^{2}P(n_i,s_i)\Big] \nonumber \\
    &=&\displaystyle\sum_{n_i=0}^{1}P(y_i|n_i)\Big[\displaystyle\sum_{s_i=1}^{2}P(n_i|s_i)P(s_i)\Big]. \label{eq:rec} \\
\end{eqnarray}
Therefore, with the help of Eqs.~\ref{eq:joint}, \ref{eq:dyn} and \ref{eq:rec} one can compute the MI described by Eq.~\ref{eq:mi} exactly for any time interval `$i$' and given initial conditions $P(s_o)$. This expression is then used for quantifying the variation in the amount of information transmitted by the sender enzyme to the receiver protein in the signaling pathway. In the main text, we have explored how this information varies with various model parameters, namely-- switching dynamics and the production rates. This gives valuable insights into the impact of allosteric regulation on the transfer of information within a signaling pathway.

\newpage
\bibliography{Allosteric_enz}